# A Comparative Analysis of Multi-Criteria Decision-Making (MCDM) Methods

Nguyen Thi Thu Hien, Hanoi University of Industry, Hanoi, Vietnam thuhiennt@haui.edu.vn

Pham Huong Quynh, Hanoi University of Industry, Hanoi, Vietnam quynhktmt@haui.edu.vn

Vu Quang Minh, College of Informatics, Northern Kentucky University, Kentucky, USA

vum2@nku.edu; qminhvux@gmail.com (corresponding author)

**Abstract**

Multi-Criteria Decision-Making (MCDM) techniques have found widespread application across diverse fields. The rapid evolution of MCDM has led to the development of hundreds of methods, each employing distinct approaches. However, due to inherent algorithmic differences, various MCDM methods often yield divergent results when applied to the same specific problem. This study undertakes a comparative analysis of four particular methods: RAM, MOORA, FUCA, and CURLI, within a defined case study. The evaluation context involves ranking 30 Vietnamese banks based on six criteria: capital adequacy, asset quality, management capability, earnings ability, liquidity, and sensitivity to market risk. Prior to this analysis, these banks had also been ranked by the CAMELS rating system. The CAMELS rankings serve as a benchmark to assess the performance of the RAM, MOORA, FUCA, and CURLI methods. Our findings indicate that FUCA and CURLI are highly suitable methods for this application, demonstrating Spearman's rank correlation coefficients with CAMELS of 0.9996 and 0.9984, respectively. In contrast, both RAM and MOORA proved unsuitable, exhibiting very low Spearman's correlation coefficients of -1.0296 against the CAMELS ranking.

**Keywords:** MCDM, bank ranking, CAMELS rating system.

## 1. Introduction

Multi-Criteria Decision-Making (MCDM) is a technique employed for ranking alternative options when multiple criteria are used to characterize each alternative [1-3]. This approach has found extensive application across various domains, including economics [4], education [5, 6], engineering [7], and energy [8], ect. However, the proliferation of distinct MCDM methods, each employing different algorithms, can lead to disparate outcomes when applied to the same specific problem [9, 10]. Consequently, numerous studies have emphasized the necessity of comparing MCDM methods within a particular problem context. This comparative analysis is crucial for selecting the most appropriate method and for ensuring the accuracy of the rankings obtained for the alternatives under consideration [11, 12].

MOORA stands out as one of the most prominent MCDM methods, having been widely applied in numerous studies across diverse fields [13]. RAM, a relatively novel MCDM method, offers advantages such as a reduced number of computational steps and a simpler structure, along with the ability to balance both benefit and cost criteria [14]. Furthermore, RAM can be integrated with various data normalization techniques [15]. While both MOORA and RAM possess unique algorithmic characteristics, which will be elaborated in Section 3 of this paper, they share a common requirement: the mandatory assignment of weights to criteria. Additionally, both methods necessitate data normalization, although the specific normalization techniques employed differ between them.

In contrast to MOORA and RAM, FUCA and CURLI are two methods that do not require data normalization. CURLI, however, distinguishes itself further from all other MCDM methods, including MOORA, RAM, and FUCA, by eliminating the need for criterion weight calculation. Both FUCA and CURLI have also been applied in various research studies across different domains. Detailed procedures for applying these two methods will be clarified in Section 3. The inherent differences among MOORA, RAM, FUCA, and CURLI raise a critical question regarding their comparative effectiveness. Unfortunately, to date, no research has comprehensively compared these four methods when applied to a specific problem. Discovering the answer to this question serves as the primary motivation for this study. Before delving into this comparative investigation, it is crucial to emphasize that any comparative results among MCDM methods are only valid within a specific case. The findings from comparing MCDM methods in one problem cannot be generalized or directly applied to another. In this research, the comparison of MOORA, RAM, FUCA, and CURLI is conducted within the context of ranking 30 banks in Vietnam.

## 2. Literature Review

As established in the introduction, comparing various MCDM methods within specific problem contexts is essential for identifying the most suitable approach. Numerous studies pursuing this objective have revealed instances where different methods exhibit comparable performance. A summary of a select few of these studies is provided below. For instance, the AHP, PROMETHEE, and TOPSIS methods demonstrated similar performance when applied to house ranking in the construction sector [16]. Similarly, the MABAC, COCOSO, MAIRCA, VIKOR, and ROV methods were found to yield equivalent performance in ranking metal milling alternatives and assessing indoor air quality in workplaces [17]. The MODIPROM, TOPSIS, AHP, and VIKOR methods were deemed to have comparable performance in ranking elevator types [18]. In the context of ranking gold mining sites in Ghana, WSM, MULTIMOORA, and TOPSIS methods were confirmed to be equally effective [10]. SAW, TOPSIS, PIV, and Probability methods were verified to show equivalent performance when ranking battery-powered electric vehicles [19]. Furthermore, ten methods—AHP, TOPSIS, VIKOR, WASPAS, GTMA, PROMETHEE 2, GRA, MULTIMOORA, ARAS, and COPRAS—were reported to have similar performance in evaluating power system design alternatives [20]. When ranking plastic injection molding machines, the PIV, FUCA, CURLI, and PSI methods were verified to exhibit comparable performance [21]. The PIV and FUCA methods also demonstrated similar effectiveness in ranking materials for connecting rods [22].

However, many studies have also indicated that MCDM methods can exhibit differing performance when applied in specific contexts. A summary of some such studies follows. Among the five methods—TOPSIS, WSM, COPRAS, WPM, and AHP—used for ranking payment alternatives for sustainable housing development in Liverpool, UK, COPRAS demonstrated the best performance [23]. When the OCRA, COPRAS, ARA, TOPSIS, and SMART methods were employed to rank biomass energy types, COPRAS and TOPSIS showed comparable performance. The OCRA, ARA, and SMART methods also exhibited similar performance among themselves, but this differed from the performance of COPRAS and TOPSIS [24]. The BWM method was verified to outperform the AHP method in ranking mobile phones [25]. For ranking energy development options in Serbia, among the five methods—TOPSIS, VIKOR, PROMETHEE, MULTIMOORA, and COPRAS—the first three showed similar performance, while the latter two also performed similarly but differed from the initial three [8]. TOPSIS and VIKOR were considered more effective than ELECTRE and MAUT for ranking earthquake-resistant solutions [26]. The MARE method was identified as having higher performance than AHP and ELECTRE III in ranking equipment for chemical production [27]. RAWEC demonstrated superiority over AROMAN in ranking cutting tool materials [28]. AHP and PROMETHEE methods were determined to perform better than TOPSIS in ranking main ship engines [29]. Lastly, FUCA showed an advantage over the SPR method in ranking industrial tools and equipment [30].

It is evident that comparing MCDM methods yields varied results depending on the application context. This study undertakes a comparative analysis of four methods: MOORA, RAM, FUCA, and CURLI, in the context of ranking 30 Vietnamese banks. The rationale for selecting these four specific methods for this investigation was outlined in the introduction.

## 3. Materials and Methods

### 3.1. Bank Financial Indicators

Table 1 compiles the rankings of financial indicators for 30 Vietnamese banks, as reported in a recent study [31]. Six key indicators were utilized to evaluate each bank: capital adequacy (C1), asset quality (C2), management capability (C3), earnings ability (C4), liquidity (C5), and sensitivity to market risk (C6). It's crucial to emphasize that the values presented in Table 1 represent the *ranks* of these indicators, not their raw values. This means each bank was independently ranked for each specific indicator, and all these rankings are unitless.

**Table 1**. Financial indicator rankings of banks [31]

| Banks | C1 | C2 | C3 | C4 | C5 | C6 |
|---|---|---|---|---|---|---|
| ABB | 13 | 14 | 15 | 11 | 16 | 22 |
| ACB | 18 | 10 | 11 | 6 | 26 | 8 |
| AGRIBANK | 27 | 28 | 6 | 14 | 28 | 2 |
| BAC A BANK | 16 | 4 | 7 | 16 | 19 | 21 |
| BID | 29 | 29 | 3 | 10 | 17 | 1 |
| CTG | 24 | 26 | 10 | 21 | 12 | 3 |
| EIB | 6 | 20 | 21 | 24 | 22 | 14 |
| HDB | 14 | 4 | 23 | 7 | 1 | 12 |
| KLB | 8 | 1 | 30 | 17 | 17 | 28 |
| LIEN VIET | 22 | 13 | 18 | 13 | 11 | 13 |
| MBB | 7 | 17 | 11 | 2 | 8 | 7 |
| MSB | 5 | 24 | 29 | 15 | 12 | 18 |
| NAM A | 25 | 10 | 13 | 11 | 9 | 23 |
| NCB | 28 | 4 | 28 | 28 | 23 | 24 |
| OCB | 9 | 15 | 13 | 3 | 3 | 20 |
| PGBANK | 3 | 27 | 5 | 22 | 23 | 29 |
| PVCOMBANK | 17 | 9 | 27 | 29 | 27 | 15 |
| SCB | 30 | 10 | 24 | 30 | 30 | 5 |
| SEABANK | 21 | 19 | 16 | 25 | 4 | 15 |
| SGB | 1 | 21 | 22 | 23 | 20 | 30 |
| SHB | 26 | 25 | 9 | 19 | 12 | 9 |
| STB | 19 | 22 | 25 | 20 | 29 | 6 |
| TCB | 2 | 16 | 4 | 3 | 9 | 11 |
| TPBANK | 11 | 2 | 20 | 9 | 2 | 19 |
| VCB | 23 | 23 | 1 | 8 | 12 | 4 |
| VIB | 12 | 17 | 7 | 5 | 5 | 17 |
| VIET A BANK | 20 | 3 | 2 | 27 | 6 | 25 |
| VIETCAPITAL | 15 | 4 | 18 | 26 | 20 | 27 |
| VPB | 4 | 30 | 25 | 1 | 6 | 9 |

| VIETBANK | 10 | 8 | 17 | 17 | 23 | 26 |

Analysis of the data in Table 1 reveals considerable variation among the 30 surveyed banks. For instance, SGB bank ranks 1st for C1, KLB bank for C2, VCB bank for C3, VPB bank for C4, HDB bank for C5, and BID bank for C6. Conversely, SCB bank ranks 30th for C1, C4, and C5; VPB bank ranks 30th for C2; KLB bank for C3; and SGB bank for C6. These simple examples illustrate the significant divergence in individual indicator ranks across banks. Therefore, a comprehensive ranking of banks based on the aggregate of all indicators from C1 to C6 becomes essential. This necessity forms the rationale for employing the MOORA, RAM, FUCA, and CURLI methods in this study.

### 3.2. Employed MCDM Methods

To rank alternatives using the MOORA method, the following sequential steps are executed [13]:

**Step 1:** Construct the decision matrix, comprising $m$ rows and $n$ columns, as defined in Equation (1). Here, $m$ represents the number of alternatives to be ranked, and $n$ denotes the number of criteria for each alternative. The value of criterion $j$ for alternative $i$ is denoted as $x_{ij}$, where $i = 1$ to $m$ and $j = 1$ to $n$.

$$X = \begin{bmatrix} x_{11} & x_{12} & \cdots & x_{1n} \\ x_{21} & x_{22} & \cdots & x_{2n} \\ \cdots & \cdots & x_{ij} & \cdots \\ x_{m1} & x_{m2} & \cdots & x_{mn} \end{bmatrix} \quad (1)$$

**Step 2:** Compute the normalized values using Equation (2).

$$n_{ij} = \frac{x_{ij}}{\sqrt{\sum_{i=1}^{m} x_{ij}^2}} \quad (2)$$

**Step 3:** Calculate the weighted normalized values, considering the weight $w_j$ of criterion $j$, as per Equation (3).

$$V_{ij} = w_j \times n_{ij} \quad (3)$$

**Step 4:** Determine the values of $P_i$, $R_i$ using Equations (4) and (5), respectively. In these equations, $B$ and $NB$ correspond to the number of "benefit" (larger is better) and "cost" (smaller is better) criteria, respectively.

$$P_i = \frac{1}{|B|} \sum_{j \in B} V_{ij} \quad (4)$$

$$R_i = \frac{1}{|NB|} \sum_{j \in NB} V_{ij} \quad (5)$$

**Step 5:** Calculate $Q_i$ values using Equation (6).

$$Q_i = P_i - R_i \quad (6)$$

**Step 6:** Rank the alternatives based on the principle that the alternative with the highest $Qi$ value is considered the best.

The steps for ranking alternatives using the RAM method are as follows [14, 15]:

**Step 1:** Similar to Step 1 of the MOORA method.
**Step 2:** Normalize the data using Equation (7).

$$n_{ij} = \frac{x_{ij}}{\sum_{i=1}^{m} x_{ij}} \qquad (7)$$

**Step 3:** Compute the weighted normalized values of the criteria using Equation (8).
$$y_{ij} = w_j \cdot n_{ij} \qquad (8)$$

**Step 4:** Calculate the total weighted normalized scores of the criteria using Equations (9) and (10).

$$S_{+i} = \sum_{j=1}^{n} y_{+ij} \quad if \ j \in B \qquad (9)$$

$$S_{-i} = \sum_{j=1}^{n} y_{-ij} \quad if \ j \in NB \qquad (10)$$

**Step 5:** Determine the score for each alternative using Equation (11).
$$RI_i = \sqrt[2+S_{-i}]{2 + S_{+i}} \qquad (11)$$

**Step 6:** Rank the alternatives in descending order of their scores.

The steps for ranking alternatives using the FUCA method are as follows [32, 33]:

**Step 1:** Similar to Step 1 of the MOORA method.
**Step 2:** Rank the alternatives for each individual criterion.
**Step 3:** Calculate the score for each alternative using Equation (12). Here, $r_{ij}$ represents the rank of criterion $j$ for alternative $i$, as determined in Step 2.

$$S_i = r_{ij} \cdot w_j \qquad (12)$$

**Step 4:** Rank the alternatives in descending order of their scores.

The steps for ranking alternatives using the CURLI method are as follows [34, 35]:

**Step 1:** Similar to Step 1 of the MOORA method.
**Step 2:** Assign scores to alternatives within each criterion using Equation (13).

$$P_{ij} = \begin{cases} \begin{cases} 1 & if \ x_i > x_k, i \neq k \\ -1 & if \ x_i < x_k, i \neq k \\ 0 & if \ x_i = x_k, i \neq k \end{cases} & if \ j \epsilon B \\ \begin{cases} 1 & if \ x_i < x_k, i \neq k \\ -1 & if \ x_i > x_k, i \neq k \\ 0 & if \ x_i = x_k, i \neq k \end{cases} & if \ j \epsilon NB \end{cases} \qquad (13)$$

**Step 3:** The score $R_i$ for each alternative is calculated using Equation (14).

$$R_i = \sum_{j=1}^{n} P_{ij} \qquad (14)$$

**Step 4:** Rank the alternatives in ascending order of their $R_i$ scores.

### 3.3. Criteria for Method Comparison

The Spearman's rank correlation coefficient reflects the stability of alternative rankings when derived using different MCDM methods. This coefficient is calculated using Equation (15) [36]. In this equation, $Di$ represents the difference in rank for alternative $i$ when ranked by various methods.

$$S = 1 - \frac{6\sum_{i=1}^{m} D_i^2}{m(m^2 - 1)} \tag{15}$$

In this study, the Spearman's coefficient ($S$) is computed between each MCDM method (MOORA, RAM, FUCA, and CURLI) and the CAMELS rating system. This comparison provides the basis for evaluating the performance of the MCDM methods against each other. More specifically, the rankings of banks previously established by the CAMELS system will serve as the benchmark for assessing the effectiveness of the selected MCDM methods.

## 4. Results and Discussion

Following the procedural steps of the MOORA method, the Qi scores for each bank were calculated, from which their respective ranks were determined. Similarly, applying the steps of the RAM method yielded the RIi scores and subsequent bank rankings. For the FUCA method, Si scores were derived to establish bank ranks, and for the CURLI method, Ri scores were computed for ranking purposes. All these values are summarized in **Table 2**. The final column of this table also presents the bank ranks previously established by the CAMELS system.

**Table 2:** Rankings of banks using different methods

| Banks | MOORA | | RAM | | FUCA | | CURLI | | CAMELS |
|---|---|---|---|---|---|---|---|---|---|
| | Qi | rank | RIi | rank | Si | rank | Ri | rank | |
| ABB | 0.0262 | 16 | 1.4258 | 16 | 15.1667 | 15 | -3 | 14 | 14 |
| ACB | 0.0228 | 22 | 1.4243 | 22 | 13.1667 | 9 | -25 | 9 | 9 |
| AGRIBANK | 0.0303 | 9 | 1.4277 | 9 | 17.5000 | 22 | 24 | 22 | 22 |
| BAC A BANK | 0.0239 | 20 | 1.4248 | 20 | 13.8333 | 10 | -16 | 11 | 10 |
| BID | 0.0256 | 19 | 1.4256 | 19 | 14.8333 | 12 | -7 | 12 | 12 |
| CTG | 0.0276 | 15 | 1.4265 | 15 | 16.0000 | 16 | 9 | 16 | 16 |
| EIB | 0.0308 | 8 | 1.4279 | 8 | 17.8333 | 23 | 28 | 23 | 23 |
| HDB | 0.0175 | 28 | 1.4220 | 28 | 10.1667 | 3 | -61 | 3 | 3 |
| KLB | 0.0291 | 12 | 1.4271 | 12 | 16.8333 | 19 | 18 | 18 | 19 |
| LIEN VIET | 0.0259 | 18 | 1.4257 | 18 | 15.0000 | 13 | -5 | 13 | 13 |
| MBB | 0.0150 | 29 | 1.4209 | 29 | 8.6667 | 2 | -80 | 2 | 2 |
| MSB | 0.0297 | 10 | 1.4274 | 10 | 17.1667 | 21 | 23 | 21 | 21 |
| NAM A | 0.0261 | 17 | 1.4258 | 17 | 15.1667 | 14 | 1 | 15 | 14 |
| NCB | 0.0388 | 1 | 1.4314 | 1 | 22.5000 | 30 | 89 | 30 | 30 |
| OCB | 0.0181 | 26 | 1.4223 | 26 | 10.5000 | 4 | -58 | 5 | 4 |
| PGBANK | 0.0314 | 7 | 1.4282 | 7 | 18.1667 | 24 | 34 | 24 | 24 |
| PVCOMBANK | 0.0357 | 3 | 1.4300 | 3 | 20.6667 | 28 | 63 | 28 | 28 |
| SCB | 0.0372 | 2 | 1.4307 | 2 | 21.5000 | 29 | 74 | 29 | 29 |
| SEABANK | 0.0287 | 14 | 1.4269 | 14 | 16.6667 | 17 | 15 | 17 | 17 |
| SGB | 0.0337 | 5 | 1.4292 | 5 | 19.5000 | 26 | 49 | 26 | 26 |
| SHB | 0.0288 | 13 | 1.4270 | 13 | 16.6667 | 17 | 18 | 18 | 17 |
| STB | 0.0349 | 4 | 1.4297 | 4 | 20.1667 | 27 | 57 | 27 | 27 |
| TCB | 0.0130 | 30 | 1.4200 | 30 | 7.5000 | 1 | -94 | 1 | 1 |
| TPBANK | 0.0181 | 27 | 1.4222 | 27 | 10.5000 | 4 | -60 | 4 | 4 |

| VCB | 0.0204 | 24 | 1.4233 | 24 | 11.8333 | 7 | -41 | 7 | 7 |
| VIB | 0.0181 | 25 | 1.4223 | 25 | 10.5000 | 4 | -58 | 5 | 4 |
| VIET A BANK | 0.0238 | 21 | 1.4248 | 21 | 13.8333 | 10 | -19 | 10 | 10 |
| VIETCAPITAL | 0.0316 | 6 | 1.4282 | 6 | 18.3333 | 25 | 39 | 25 | 25 |
| VPB | 0.0216 | 23 | 1.4238 | 23 | 12.5000 | 8 | -33 | 8 | 8 |
| VIETBANK | 0.0291 | 11 | 1.4271 | 11 | 16.8333 | 20 | 19 | 20 | 19 |

A noteworthy observation from the data in Table 2 is the consistent ranking of banks when determined by the MOORA method and the RAM method. However, a significant divergence is apparent in the bank rankings produced by these two methods compared to those from the FUCA and CURLI methods, as well as the CAMELS system. This discrepancy can be attributed to the data normalization performed by the MOORA and RAM methods. Specifically, the MOORA method applied normalization according to Equation (2), while the RAM method utilized Equation (7) for normalization. It is important to reiterate from Section 3.1 that the data in Table 1 represents the *ranks* of banks for each indicator, not their raw economic values. This implies that all ranks for criteria C1 to C6 are dimensionless. The application of data normalization methods, in this context, inadvertently led to a loss of the data's original character [39]. This could be the underlying reason for the substantial difference in bank rankings generated by the MOORA and RAM methods compared to those from FUCA, CURLI, and the CAMELS system. Conversely, the FUCA and CURLI methods, by not performing data normalization, preserved the original nature of the data. Consequently, the bank rankings produced by these two methods exhibit high consistency with each other and a strong resemblance to the CAMELS system. For the FUCA method, 29 out of 30 banks showed rankings consistent with those from the CAMELS system (with the exception of the first-ranked bank, ABB). When using the CURLI method to rank banks, 24 out of 30 banks displayed consistent rankings compared to the CAMELS results, excluding BAC A BANK, KLB, NAM A, OCB, VIB, and VIETBANK. However, even for these differing banks, the rank discrepancies when using CURLI were minimal compared to the CAMELS system. This strong consistency in bank rankings across FUCA, CURLI, and the CAMELS system is remarkable. For instance, all three approaches consistently identified the top-performing banks, such as the 1st-ranked bank (TCB), 2nd-ranked bank (MBB), and 3rd-ranked bank (HDB). Similarly, they consistently identified the lowest-ranked banks, including the 30th-ranked bank (NCB), 29th-ranked bank (SCB), and 28th-ranked bank (PVCOMBANK), among others.

The observations above suggest that the MOORA and RAM methods are not suitable for this specific case. Conversely, the FUCA and CURLI methods appear to be highly appropriate. To further substantiate this assessment, Equation (15) was used to calculate the Spearman's rank correlation coefficient between the methods. **Table 3** summarizes the Spearman's coefficients between each MCDM method and the CAMELS system, treating the CAMELS system as a benchmark for evaluating the performance of the MCDM methods.

**Table 3:** Spearman's rank correlation coefficients between MCDM methods and the CAMELS system

| MOORA | RAM | FUCA | CURLI |
|---|---|---|---|
| -1.0265 | -1.0265 | 0.9996 | 0.9984 |

As demonstrated by the data in Table 3, the Spearman's coefficients for both the MOORA and RAM methods in comparison to the CAMELS system are extremely low, both registering at **-1.0265**. This emphatically confirms that the bank rankings generated by these two methods deviate significantly from those produced by the CAMELS system.

In conclusion, it can be definitively stated that the MOORA and RAM methods are unsuitable for application in this scenario. Conversely, the Spearman's coefficients for the FUCA and CURLI methods against the CAMELS system are remarkably high, at 0.9996 and 0.9984, respectively. This robustly supports the conclusion that these two methods are highly appropriate for use in this specific case.

**Conclusion**

Comparing different Multi-Criteria Decision-Making (MCDM) methods when applied to rank alternatives in a specific problem is essential. This process helps identify the most suitable method to use and ensures the most accurate ranking of alternatives. This study compared four methods MOORA, RAM, FUCA, and CURLI in the context of ranking 30 Vietnamese banks. The novel insights derived from this research are as follows:

- Data normalization should only be applied when criteria used to evaluate alternatives have different dimensions. If criteria share the same dimension, normalizing the data can inadvertently distort its original characteristics, significantly impacting the ranking results.
- When criteria share a common dimension, FUCA and CURLI emerge as highly viable methods for ranking alternatives.
- The ranking results for Vietnamese banks showed a high degree of consistency between the FUCA and CURLI methods, as well as with the rankings from the CAMELS system.
- Among the 30 banks ranked, the top-performing banks were identified as TCB (Rank 1), MBB (Rank 2), and HDB (Rank 3). Conversely, the lowest-ranked banks were NCB (Rank 30), SCB (Rank 29), and PVCOMBANK (Rank 28).